\documentclass[11pt]{article}

\newcommand{\text}{\rm}

\begin{document}

\title{\textbf{The anomalous dimension of the gluon-ghost mass operator in
Yang-Mills theory }}
\author{D. Dudal\thanks{%
Research Assistant of The Fund For Scientific Research-Flanders,
Belgium.},\
 H. Verschelde\thanks{%
david.dudal@ugent.be, henri.verschelde@ugent.be} \\
{\small {\textit{Ghent University }}}\\
{\small {\textit{Department of Mathematical Physics and Astronomy,
Krijgslaan 281-S9, }}}\\
{\small {\textit{B-9000 Gent, Belgium}}} \and V.E.R. Lemes, M.S.
Sarandy, R.
Sobreiro, S.P. Sorella\thanks{%
vitor@dft.if.uerj.br, sarandy@dft.if.uerj.br,
sobreiro@dft.if.uerj.br,
sorella@uerj.br} \\
{\small {\textit{UERJ - Universidade do Estado do Rio de Janeiro,}}} \\
{\small {\textit{\ Rua S\~{a}o Francisco Xavier 524, 20550-013
Maracan\~{a},
}}} {\small {\textit{Rio de Janeiro, Brazil.}}} \and M. Picariello\thanks{%
marco.picariello@mi.infn.it } \\
{\small {\textit{Universit\'{a} degli Studi di Milano, via Celoria
16,
I-20133, Milano, Italy }}}\\
{\small {\textit{and INFN\ Milano, Italy}}}
\and J.A. Gracey\thanks{jag@amtp.liv.ac.uk } \\
{\small {\textit{Theoretical Physics Division}}} \\
{\small {\textit{Department of Mathematical Sciences}}} \\
{\small {\textit{University of Liverpool }}} \\
{\small {\textit{P.O. Box 147, Liverpool, L69 3BX, United
Kingdom}}}}
\date{}
\maketitle

\vspace{-13cm} \hfill LTH--578 \vspace{13cm}


\begin{abstract}
The local composite gluon-ghost operator $\left( \frac{1}{2}A^{a\mu }A_{\mu
}^{a}+\alpha \overline{c}^{a}c^{a}\right) $ is analysed in the framework of
the algebraic renormalization in $SU(N)$ Yang-Mills theories in the Landau,
Curci-Ferrari and maximal abelian gauges. We show, to all orders of
perturbation theory, that this operator is multiplicatively renormalizable.
Furthermore, its anomalous dimension is not an independent parameter of the
theory, being given by a general expression valid in all these gauges. We
also verify the relations we obtain for the operator anomalous dimensions by
explicit 3-loop calculations in the $\overline{\mbox{MS}}$ scheme for the
Curci-Ferrari gauge.
\end{abstract}

\newpage

\section{Introduction}

Vacuum condensates are believed to play an important role in the
understanding of the nonperturbative dynamics of Yang-Mills theories. In
particular, much effort has been devoted to the study of condensates of
dimension two built up with gluons and ghosts. For instance, the relevance
of the pure gluon condensate $\left\langle A^{a\mu }A_{\mu
}^{a}\right\rangle $ in the Landau gauge has been discussed from lattice
simulations \cite{bou} as well as from a phenomenological point of view \cite
{gz}. That the operator $A^{2}_\mu$ has a special meaning in the Landau
gauge follows by observing that, due to the transversality condition $%
\partial _{\mu }A^{a\mu }=0$, the integrated operator $(VT)^{-1}\int
d^{4}x\,A_{\mu }^{a}A^{a\mu }\,\,$is gauge invariant, with $VT$ denoting the
space-time volume. An effective potential for $\left\langle A^{a\mu }A_{\mu
}^{a}\right\rangle $ has been constructed in \cite{vland}, showing that the
vacuum of Yang-Mills favors a nonvanishing value for this condensate, which
gives rise to a dynamical mass generation for the gluons.

The operator $A^{2}_\mu$ in the Landau gauge can be generalized to other
gauges such as the Curci-Ferrari gauge and maximal abelian gauge, (MAG).
Indeed, as was shown in \cite{ope}, the mixed gluon-ghost operator\footnote{%
In the case of the maximal abelian gauge, the color index $a$ runs only over
the $N(N-1)$ off-diagonal components.} $\mathcal{O}$ $=$ $\left( \frac{1}{2}%
A^{a\mu }A_{\mu }^{a}+\alpha \overline{c}^{a}c^{a}\right) $ turns out to be
BRST\ invariant on-shell, where $\alpha$ is the gauge parameter. Also, the
Curci-Ferrari gauge has the Landau gauge, $\alpha =0$, as a special case.
Thus, the gluon-ghost condensate $\left( \frac{1}{2}\left\langle A^{a\mu
}A_{\mu }^{a}\right\rangle +\alpha \left\langle \overline{c}%
^{a}c^{a}\right\rangle \right) $ might be suitable for the description of
dynamical mass generation in these gauges. Recently, the effective potential
for this condensate in the Curci-Ferrari gauge has been constructed in \cite
{cfep} by combining the algebraic renormalization \cite{book} with the local
composite operators technique \cite{vland,vlco}, resulting in a dynamical
mass generation. In this formalism, an essential step is the
renormalizability of the local composite operator related to the condensate,
which is fundamental to obtaining its anomalous dimension. It is worth
mentioning that the anomalous dimension of the gluon-ghost operator $%
\mathcal{O}$ $=$ $\left( \frac{1}{2}A^{a\mu }A_{\mu }^{a}+\alpha \overline{c}%
^{a}c^{a}\right) $ in the Curci-Ferrari gauge, and thus of the gluon
operator $A^{2}_\mu$ in the Landau gauge, has been computed to three loops
in the $\overline{\mbox{MS}}$ renormalization scheme, \cite{gracey}. In
addition, it has been proven \cite{dsv} by using BRST\ Ward identities that
the anomalous dimension $\gamma _{A^{2}_\mu}(a)$ of the operator $A^{2}_\mu$
in the Landau gauge is not an independent parameter, being expressed as a
combination of the gauge beta function, $\beta(a)$, and of the anomalous
dimension, $\gamma_{A}(a)$, of the gauge field, according to the relation
\begin{equation}
\gamma _{A^{2}_\mu}(a) ~=~ -~ \left( \frac{\beta (a)}{a} ~+~ \gamma
_{A}(a)\right) ,\,\,\,\,\,\,\,\,\,\,\,a=\frac{g^{2}}{16\pi ^{2}}\,\,\,.
\label{ada2}
\end{equation}
The aim of this paper is to extend the analysis of \cite{dsv} to the
Curci-Ferrari and maximal abelian gauges. We shall prove that the operator $%
\left( \frac{1}{2}A^{a\mu }A_{\mu }^{a}+\alpha \overline{c}^{a}c^{a}\right)$
is multiplicatively renormalizable to all orders of perturbation theory.
Furthermore, as in the case of the Landau gauge, its anomalous dimension $%
\gamma_{\mathcal{O}}(a)$ is not an independent parameter of the theory,
being given in fact by a general relationship valid in the Landau,
Curci-Ferrari and maximal abelian gauges, which is
\begin{equation}
\gamma_{\mathcal{O}}(a) ~=~ -~ 2\left( \gamma _{c}(a) ~+~ \gamma
_{gc^{2}}(a) \right) ~,  \label{go}
\end{equation}
where $\gamma_{c}(a)$, $\gamma_{gc^{2}}(a)$ are the anomalous dimensions of
the Faddeev-Popov ghost $c^{a}$ and of the composite operator $\frac{1}{2}%
gf^{abc}c^{b}c^{c}$ corresponding to the BRST variation of $c^{a}$. In other
words $sc^{a}$~$=$~$\frac{1}{2}gf^{abc}c^{b}c^{c}$ where $s$ is the BRST
operator.

The paper is organized as follows. In section 2, the renormalization of the
dimension two operator $\left( \frac{1}{2}A^{a\mu }A_{\mu }^{a}+\alpha
\overline{c}^{a}c^{a}\right)$ is considered in detail, by taking the
Curci-Ferrari gauge as an example and the relationship (\ref{go}) is
derived. In section 3, we verify the relation we obtain between the
anomalous dimensions explicitly at three loops in the Curci-Ferrari gauge in
the $\overline{\mbox{MS}}$ scheme. Section 4 is devoted to the Landau gauge,
showing that the expression (\ref{go}) reduces to that of (\ref{ada2}). In
section 5 we shall analyse the maximal abelian gauge, where the results
established in \cite{ell} for the case of $SU(2)$ will be recovered.
Finally, in section 6 we present our conclusions.

\section{The gluon-ghost operator in Yang-Mills theories in the
Curci-Ferrari gauge}

\subsection{The Curci-Ferrari action}

We begin by reviewing the quantization of pure $SU(N)$ Yang-Mills in the
Curci-Ferrari gauge. The pure Yang-Mills action is given by
\begin{equation}
S_{\mathrm{YM}} ~=~ -~ \frac{1}{4}\int d^{4}x \,
F^{a\,\mu\nu}F_{\mu\nu}^{a}\;,  \label{symp}
\end{equation}
with $F_{\mu \nu }^{a}$ $=$ $\partial _{\mu }A_{\nu}^{a}-\partial_{\nu
}A_{\mu }^{a}+gf^{abc}A_{\mu }^{b}A_{\nu }^{c}$. The so called Curci-Ferrari
gauge, \cite{obt,cf}, is defined by the following gauge fixing term
\begin{eqnarray}
S_{\mathrm{gf}}\; &=&\int d^{4}x\left( b^{a}\partial _{\mu }A^{\mu a}+\frac{%
\alpha }{2}b^{a}b^{a}+\,\overline{c}^{a}\partial ^{\mu }\left( D_{\mu
}c\right) ^{a}\right.  \nonumber \\
&&\left. ~~~~~~~~~~-~ \frac{\alpha }{2}gf^{abc}b^{a}\overline{c}^{b}c^{c} -%
\frac{\alpha}{8}g^{2}f^{abc}\overline{c}^{a}\overline{c}%
^{b}f^{cmn}c^{m}c^{n}\right) \;,  \label{sgf}
\end{eqnarray}
with
\begin{equation}
\left( D_{\mu}c\right)^{a} ~=~ \partial _{\mu }c^{a} ~+~ gf^{abc}A_{\mu}^{b}
c^{c}\;.  \label{cder}
\end{equation}
In order to analyse the renormalization of the operator $\left( \frac{1}{2}%
A^{a\mu }A_{\mu }^{a}+\alpha \overline{c}^{a}c^{a}\right)$, we have to
introduce it in the action by means of a set of external sources. Following
\cite{dsv}, it turns out that three external sources $J$, $\eta^{\mu }$ and $%
\tau ^{\mu }$ are required, so that
\begin{eqnarray}
S_{\mathrm{J}} &=& \int d^{4}x\left( J\left( \frac{1}{2}A^{a\mu }A_{\mu
}^{a}+\alpha \overline{c}^{a}c^{a}\right) +\,\frac{\xi }{2}J^{2}-\eta ^{\mu
}A_{\mu }^{a}c^{a}-\tau ^{\mu }\left( \partial _{\mu }c^{a}\right)
c^{a}\right.  \nonumber \\
&&\left. \,\,\,\,\,\,\,\,\,\,\,\,\,\,\,\,\,\,\,\,-~ \frac{g}{2}\tau^{\mu}
f^{abc}A_{\mu }^{a}c^{b}c^{c}\right)  \label{sop}
\end{eqnarray}
where $\xi$ is a dimensionless parameter, accounting for the divergences
present in the vacuum Green function $\left\langle \left( \frac{1}{2}A^{a\mu
}A_{\mu }^{a}+\alpha \overline{c}^{a}c^{a}\right) _{x}\left( \frac{1}{2}%
A^{a\mu }A_{\mu }^{a}+\alpha \overline{c}^{a}c^{a}\right) _{y}\right\rangle $%
, which are proportional to $J^{2}$ \cite{cfep}. The action $\left( S_{%
\mathrm{YM}}+S_{\mathrm{gf}}+S_{\mathrm{J}}\right) $ is invariant under the
BRST\ transformations, which read
\begin{eqnarray}
&&sA_{\mu }^{a} ~=~ -\left( D_{\mu }c\right)^{a}\,\,,\,\,sc^{a} ~=~ \frac{g}{%
2} f^{abc}c^{b}c^{c}\;,  \nonumber \\
&&s\overline{c}^{a} ~=~ b^{a}\,\,,\hspace{1.6cm}sb^{a} ~=~ -Jc^{a}\;,
\nonumber \\
&&s\tau ^{\mu } ~=~ -\eta ^{\mu }\,\,,\hspace{1.2cm}s\eta^{\mu} ~=~
\partial^{\mu}J\;,  \nonumber \\
&&sJ ~=~ 0\;.  \label{s}
\end{eqnarray}
Also, due to the introduction of the external sources $J$, $\eta ^{\mu }$
and $\tau^{\mu}$ it follows that the operator $s$ is not strictly nilpotent,
namely
\begin{eqnarray}
&&s^{2}\Phi ~=~ 0\;,\,\,\,\;(\Phi =A^{a\mu },\,c^{a},\,J,\,\eta ^{\mu })\;,
\nonumber \\
&&s^{2}\overline{c}^{a} ~=~ -~ Jc^{a}\,,  \nonumber \\
&&s^{2}b^{a} ~=~ -~ J\frac{g}{2}f^{abc}c^{b}c^{c}\,,  \nonumber \\
&&s^{2}\tau ^{\mu } ~=~ -~ \partial ^{\mu }J\;.  \label{s2}
\end{eqnarray}
Therefore, setting
\begin{equation}
s^{2} ~\equiv~ \delta \;,  \label{d}
\end{equation}
we have $\delta \left( S_{\mathrm{YM}}+S_{\mathrm{gf}}+S_{\mathrm{J}}\right)
=0.$ The operator $\delta $ is related to a global $SL(2,R)$ symmetry \cite
{fd}, which is known to be present in the Landau, Curci-Ferrari and maximal
abelian gauges, \cite{sl2r}. Finally, in order to express the BRST\ and $%
\delta$ invariances in a functional way, we introduce the external action
\begin{eqnarray}
S_{\mathrm{ext}} &=&\int d^{4}x\left( \Omega ^{a\mu }\,sA_{\mu
}^{a}\,+\,L^{a}\,sc^{a}\right)  \label{sext} \\
&=&\int d^{4}x\left( -~ \Omega ^{a\mu }\,\left( D_{\mu }c\right)
^{a}+\,L^{a}\,\frac{g}{2}f^{abc}c^{b}c^{c}\right) ,  \nonumber
\end{eqnarray}
where $\Omega^{a\mu}$ and $L^{a}$ are external sources invariant under both
BRST and $\delta$ transformations, coupled to the nonlinear variations of
the fields $A_{\mu }^{a}$ and $c^{a}$. It is worth noting that the source $%
L^{a}$ couples to the composite operator $\frac{g}{2}f^{abc}c^{b}c^{c}$,
thus defining its renormalization properties. It is easy to check that the
complete classical action
\begin{equation}
\Sigma ~=~ S_{\mathrm{YM}} ~+~ S_{\mathrm{gf}} ~+~ S_{\mathrm{J}} ~+~ S_{%
\mathrm{ext}}  \label{ca}
\end{equation}
is invariant under BRST and $\delta$ transformations
\begin{equation}
s\Sigma ~=~ 0\,\,\,,\,\,\,\,\delta \Sigma ~=~ 0\;.  \label{inv}
\end{equation}
When translated into functional form, the BRST and the $\delta$ invariances
give rise to the following Ward identities for the complete action $\Sigma$,
namely

\begin{itemize}
\item  the Slavnov-Taylor identity
\end{itemize}

\begin{equation}
\mathcal{S}(\Sigma ) ~=~ 0\;,  \label{sti}
\end{equation}

\begin{itemize}
\item  with
\begin{equation}
\mathcal{S}(\Sigma )=\int d^{4}x\,\left( \frac{\delta \Sigma }{\delta \Omega
^{a\mu }}\frac{\delta \Sigma }{\delta A_{\mu }^{a}}+\frac{\delta \Sigma }{%
\delta L^{a}}\frac{\delta \Sigma }{\delta c^{a}}+b^{a}\frac{\delta \Sigma }{%
\delta \overline{c}^{a}}-Jc^{a}\frac{\delta \Sigma }{\delta b^{a}}+\partial
^{\mu }J\frac{\delta \Sigma }{\delta \eta ^{\mu }}-\eta ^{\mu }\frac{\delta
\Sigma }{\delta \tau ^{\mu }}\right) \;  \label{std}
\end{equation}

\item  The $\delta $ Ward identity
\begin{equation}
\mathcal{W}\left( \Sigma \right) ~=~0\;,  \label{wwi}
\end{equation}
with
\begin{equation}
\mathcal{W}(\Sigma )=\int d^{4}x\,\left( -Jc^{a}\frac{\delta \Sigma }{\delta
\overline{c}^{a}}-J\frac{\delta \Sigma }{\delta L^{a}}\frac{\delta \Sigma }{%
\delta b^{a}}-\partial ^{\mu }J\frac{\delta \Sigma }{\delta \tau ^{\mu }}%
\right) \;.  \label{wwd}
\end{equation}
\end{itemize}

\subsection{The invariant counterterm and the renormalization constants}

We are now ready to analyse the structure of the most general local
counterterm compatible with the identities (\ref{sti}) and (\ref{wwi}). Let
us begin by displaying the quantum numbers of all fields and sources
\begin{equation}
\begin{tabular}{|l|l|l|l|l|l|l|l|l|l|}
\hline
& $A_{\mu }^{a}$ & $c^{a}$ & $\overline{c}^{a}$ & $b^{a}$ & $L^{a}$ & $%
\Omega _{\mu }^{a}$ & $J$ & $\eta ^{\mu }$ & $\tau ^{\mu }$ \\ \hline
Gh. number & $0$ & $1$ & $-1$ & $0$ & $-2$ & $-1$ & $0$ & $-1$ & $-2$ \\
\hline
Dimension & $1$ & $0$ & $2$ & $2$ & $4$ & $3$ & $2$ & $3$ & $3$ \\ \hline
\end{tabular}
\label{t1}
\end{equation}
In order to characterize the most general invariant counterterm which can be
freely added to all orders of perturbation theory, we perturb the classical
action $\Sigma $ by adding an arbitrary integrated local polynomial $\Sigma
^{\mathrm{count}}$ in the fields and external sources of dimension bounded
by four and with zero ghost number, and we require that the perturbed action
$(\Sigma +\varepsilon \Sigma ^{\mathrm{count}})$ satisfies the same Ward
identities and constraints as $\Sigma $ to first order in the perturbation
parameter $\varepsilon$, which are
\begin{eqnarray}
\mathcal{S}(\Sigma +\varepsilon \Sigma ^{\mathrm{count}})
&=&0~+~O(\varepsilon ^{2})\;,  \nonumber \\
\mathcal{W}\left( \Sigma +\varepsilon \Sigma ^{\mathrm{count}}\right)
&=&0~+~O(\varepsilon ^{2})\;,  \label{eps}
\end{eqnarray}
This amounts to imposing the following conditions on $\Sigma^{\mathrm{count}%
} $%
\begin{equation}
\mathcal{B}_{\Sigma }\Sigma ^{\mathrm{count}} ~=~ 0\;,  \label{stt}
\end{equation}
with
\begin{eqnarray}
\mathcal{B}_{\Sigma } &=&\int d^{4}x\left( \frac{\delta \Sigma }{\delta
A_{\mu }^{a}}\frac{\delta }{\delta \Omega ^{a\mu }}+\frac{\delta \Sigma }{%
\delta \Omega ^{a\mu }}\frac{\delta }{\delta A_{\mu }^{a}}+\frac{\delta
\Sigma }{\delta L^{a}}\frac{\delta }{\delta c^{a}}+\frac{\delta \Sigma }{%
\delta c^{a}}\frac{\delta }{\delta L^{a}}\right.  \label{bs} \\
&&\left. \;\;\;\;\;\;\;+~b^{a}\frac{\delta }{\delta \overline{c}^{a}}%
+\partial _{\mu }J\frac{\delta }{\delta \eta _{\mu }}+\eta ^{\mu }\frac{%
\delta }{\delta \tau ^{\mu }}-Jc^{a}\frac{\delta }{\delta b^{a}}\right) \;,
\nonumber
\end{eqnarray}
and
\begin{equation}
\int d^{4}x\,\left( -Jc^{a}\frac{\delta \Sigma ^{\mathrm{count}}}{\delta
\overline{c}^{a}}-J\frac{\delta \Sigma }{\delta L^{a}}\frac{\delta \Sigma ^{%
\mathrm{count}}}{\delta b^{a}}-J\frac{\delta \Sigma }{\delta b^{a}}\frac{%
\delta \Sigma ^{\mathrm{count}}}{\delta L^{a}}-\partial ^{\mu }J\frac{\delta
\Sigma ^{\mathrm{count}}}{\delta \tau ^{\mu }}\right) ~=~ 0\;.  \label{st2}
\end{equation}
Proceeding as in \cite{dsv}, it turns out that the most general local
invariant counterterm compatible with the Ward identities (\ref{sti}) and (%
\ref{wwi}) contains six independent parameters denoted by $\sigma$, $a_{1}$,
$a_{2}$, $a_{3}$, $a_{4}$ and $a_{5}$, and is given by
\begin{eqnarray}
\Sigma ^{\mathrm{count}} &=&\int d^{4}x\,\left( \frac{\sigma }{4}F^{a\mu \nu
}F_{\mu \nu }^{a}+(a_{3}-a_{4})(D_{\mu }F^{\mu \nu })^{a}A_{\nu }^{a}+\frac{%
a_{1}}{2}b^{a}b^{a}+a_{2}b^{a}\partial ^{\mu }A_{\mu }^{a}\right.  \nonumber
\\
&&\,\,\,\,\,\,\,\,\,\,\,\,\,+~(a_{2}-a_{3})\overline{c}^{a}\partial
^{2}c^{a}+(a_{4}-a_{2})gf^{abc}\overline{c}^{a}\partial ^{\mu }\left(
c^{b}A_{\mu }^{c}\right)  \nonumber \\
&&\,\,\,\,\,\,\,\,\,\,\,\,\,+~\frac{(\alpha a_{4}-a_{1})}{2}gf^{abc}b^{a}%
\overline{c}^{b}c^{c}+(\alpha a_{4}-\frac{a_{1}}{2})\frac{g^{2}}{4}f^{abc}%
\overline{c}^{a}\overline{c}^{b}f^{cmn}c^{m}c^{n}  \nonumber \\
&&\,\,\,\,\,\,\,\,\,\,\,\,\,+~a_{3}\Omega ^{a\mu }\partial _{\mu
}c^{a}+a_{4}gf^{abc}\Omega ^{a\mu }A_{\mu }^{b}c^{c}-\frac{a_{4}}{2}%
L^{a}gf^{abc}c^{b}c^{c}  \nonumber \\
&&\,\,\,\,\,\,\,\,\,\,\,\,\,+~\frac{(a_{2}+a_{3})}{2}JA^{a\mu }A_{\mu
}^{a}+a_{1}J\overline{c}^{a}c^{a}+a_{5}\frac{\xi }{2}J^{2}-a_{2}\eta ^{\mu
}A_{\mu }^{a}c^{a}  \nonumber \\
&&\,\,\,\,\,\,\,\,\,\,\,\,\,\left. +~(a_{2}-a_{3})\tau ^{\mu }c^{a}\partial
_{\mu }c^{a}+(a_{4}-a_{2})\frac{g}{2}\tau ^{\mu }f^{abc}A_{\mu
}^{a}c^{b}c^{c}\right) ~.  \label{ic}
\end{eqnarray}
It therefore remains to discuss the stability of the classical action. In
other words to check that $\Sigma^{\mathrm{count}}$ can be reabsorbed in the
classical action $\Sigma $ by means of a multiplicative renormalization of
the coupling constant $g$, the parameters $\alpha $ and $\xi $, the fields $%
\left\{ \phi =A,c,\overline{c},b\right\} $ and the sources $\left\{ \Phi
=J,\eta ,\tau ,L,\Omega \right\} $, namely
\begin{equation}
\Sigma (g,\xi ,\alpha ,\phi ,\Phi )+\varepsilon \Sigma ^{\mathrm{count}%
}=\Sigma (g_{0},\xi _{0},\alpha _{0},\phi _{0},\Phi _{0})+O(\varepsilon
^{2})\;,  \label{stab}
\end{equation}
with the bare fields and parameters defined as
\begin{eqnarray}
A_{0\mu }^{a} &=&Z_{A}^{1/2}A_{\mu }^{a}\,\,\,,\,\Omega _{0\mu }^{a} ~=~
Z_{\Omega }\Omega _{\mu }^{a}\,\,\,,\,\,\,\tau _{0\mu } ~=~ Z_{\tau
}\tau_{\mu }  \nonumber \\
c_{0}^{a} &=&Z_{c}^{1/2}c^{a}\,\,\,\,\,,\,\,\,L_{0}^{a} ~=~
Z_{L}L^{a}\,\,\,\,\,,\,\,\,g_{0} ~=~ Z_{g}g\,\,\,,  \nonumber \\
\overline{c}_{0}^{a} &=&Z_{\overline{c}}^{1/2}\overline{c}%
^{a}\,\,\,\,\,,\,\,\,J_{0} ~=~ Z_{J}J\,\,\,\,\,\,\,\,\,,\,\,\alpha _{0} ~=~
Z_{\alpha }\alpha \,,  \nonumber \\
b_{0}^{a} &=&Z_{b}^{1/2}b^{a}\,\,\,\,,\,\,\,\,\eta _{0\mu } ~=~ Z_{\eta
}\eta _{\mu }\,\,\,\,\,\,,\,\,\,\xi _{0} ~=~ Z_{\xi }\xi \,.  \label{rec}
\end{eqnarray}
The parameters $\sigma$, $a_{1}$, $a_{2}$, $a_{3}$, $a_{4}$ and $a_{5}$ turn
out to be related to the renormalization of the gauge coupling constant $g$,
of the gauge parameter $\alpha $, and of $L^{a}$, $c^{a}$, $A_{\mu }^{a} $,
and $\xi $ respectively, according to
\begin{eqnarray}
\,Z_{g} &=&1+\varepsilon \frac{\sigma }{2}\,\,\,\,,\,\,  \nonumber \\
Z_{\alpha } &=&1+\,\varepsilon \left( \frac{a_{1}}{\alpha }-\sigma
-2a_{2}+2a_{3}-2a_{4}\right) \,,\,\,\,  \nonumber \\
Z_{L} &=&1+\varepsilon \left( -a_{2}-\frac{\sigma }{2}+a_{3}-a_{4}\right) \,,
\nonumber \\
Z_{c}^{1/2} &=&1+\varepsilon \left( \frac{-a_{3}+a_{2}}{2}\right) \,\,,
\nonumber \\
\,\,\,\,\,Z_{A}^{1/2} &=&1+\varepsilon \left( -\frac{\sigma }{2}%
+a_{3}-a_{4}\right) ,  \nonumber \\
Z_{\xi } &=&1+\varepsilon \left( a_{5}-2\sigma -2a_{2}+2a_{3}-4a_{4}\right)
~.  \label{zetas}
\end{eqnarray}
Concerning the other fields and the sources $\Omega_{\mu}^{a}$, $\eta_{\mu}$%
, and $\tau _{\mu }$, it can be verified that they are renormalized as
\begin{eqnarray}
Z_{\overline{c}}\,
&=&Z_{c}\,\,\,,\,\,\,\,\,\,\,\,\,\;\;\;\;\;\;\,\,\,\,\,\,\,Z_{b}^{1/2} ~=~
Z_{L}^{-1}\,\,\,\,,\,\,\,\,\,Z_{\Omega } ~=~ Z_{L}\,Z_{A}^{-1/2}Z_{c}^{1/2}
\nonumber \\
Z_{\eta } &=&Z_{L}^{-1}Z_{c}^{-1/2}\,\,,\,\,\,\,\,\,\,\,\,\,Z_{\tau } ~=~ 1~.
\label{oqf}
\end{eqnarray}
Finally, for the source $J$, one has
\begin{equation}
Z_{J} ~=~ Z_{L}^{-2}\,Z_{c}^{-1},  \label{zj}
\end{equation}
from which it follows that
\begin{equation}
\gamma_{\mathcal{O}}(a) ~=~ -~ 2\left( \gamma _{c}(a) ~+~ \gamma_{gc^{2}}(a)
\right) \;,  \label{gj}
\end{equation}
where $\gamma_{c}(a)$ and $\gamma_{gc^{2}}(a)$ are the anomalous dimensions
of the Faddeev-Popov ghost $c^{a}$ and of the composite operator $\frac{g}{2}%
f^{abc}c^{b}c^{c}$, defined as
\begin{eqnarray}
\gamma_{c}(a) &=& \mu \partial_{\mu }\ln Z_{c}^{1/2} ~~~~~
\gamma_{gc^{2}}(a) ~=~ \mu \partial _{\mu }\ln Z_{L} ~~~~~ \gamma_{\mathcal{O%
}}(a) ~=~ \mu \partial _{\mu }\ln Z_{J}  \nonumber \\
\frac{\beta (a)}{a} &=& \mu \partial _{\mu }\ln Z_{g}^{-1} ~~~~~
\gamma_{\alpha}(a) ~=~ \mu \partial _{\mu }\ln Z_{\alpha }^{-1}  \label{glc}
\end{eqnarray}
where $\mu$ is the renormalization scale.

Therefore, we have provided a purely algebraic proof of the multiplicative
renormalizability of the gluon-ghost operator to all orders of perturbation
theory. In particular, we have been able to show, as explicitly exhibited in
(\ref{gj}), that the anomalous dimension of $\left( \frac{1}{2} A^{a\mu}
A_{\mu}^{a}+\alpha \overline{c}^{a}c^{a}\right)$ is not an independent
parameter of the theory, being given by a combination of the anomalous
dimensions $\gamma_{c}(a)$ and $\gamma_{gc^{2}}(a)$. It is worth mentioning
that it has also been proven, \cite{Dudal:2003dp}, for the Curci-Ferrari
gauge that the anomalous dimension of the ghost operators $\frac{g}{2}%
f^{abc}c^{b}c^{c}$, $\frac{g}{2}f^{abc}\overline{c}^{b}c^{c}$ and $\frac{g}{2%
}f^{abc}\overline{c}^{b}\overline{c}^{c}$ are the same.

Although we did not consider matter fields in the previous analysis, it can
be checked that the renormalizability of $\mathcal{O}$ and the relation (\ref
{gj}) remain unchanged if matter fields are included.

\section{Three loop verification}

In this section, we will explicitly verify the relation (\ref{gj}) up to
three loop order in the Curci-Ferrari gauge in a particular renormalization
scheme, $\overline{\mbox{MS}}$. The values for the $\beta$-function and the
anomalous dimensions of the gluon, ghost, the operator $\mathcal{O}$ and the
gauge parameter $\alpha$ have already been calculated in the presence of
matter fields in \cite{gracey}. For completeness we note that for an
arbitrary colour group these are
\begin{eqnarray}  \label{alphar}
\beta (a) &=&-~\left[ \frac{11}{3}C_{A}-\frac{4}{3}T_{F}N_{\!f}\right]
a^{2}~-~\left[ \frac{34}{3}C_{A}^{2}-4C_{F}T_{F}N_{\!f}-\frac{20}{3}%
C_{A}T_{F}N_{\!f}\right] a^{3}  \nonumber \\
&&+~\left[
2830C_{A}^{2}T_{F}N_{\!f}-2857C_{A}^{3}+1230C_{A}C_{F}T_{F}N_{%
\!f}-316C_{A}T_{F}^{2}N_{\!f}^{2}\right.  \nonumber \\
&&\left. ~~~~~-~108C_{F}^{2}T_{F}N_{\!f}-264C_{F}T_{F}^{2}N_{\!f}^{2}\right]
\frac{a^{4}}{54}~+~O(a^{5})~\;,  \label{beta} \\
\gamma _{A}(a) &=&\left[ (3\alpha -13)C_{A}+8T_{F}N_{\!f}\right] \frac{a}{6}
\nonumber  \label{gA} \\
&&+~\left[ \left( \alpha ^{2}+11\alpha -59\right)
C_{A}^{2}+40C_{A}T_{F}N_{\!f}+32C_{F}T_{F}N_{\!f}\right] \frac{a^{2}}{8}
\nonumber \\
&&+~\left[ \left( 54\alpha ^{3}+909\alpha ^{2}+(6012+864\zeta (3))\alpha
+648\zeta (3)-39860\right) C_{A}^{3}\right.  \nonumber \\
&&\left. ~~~~~-~\left( 2304\alpha +20736\zeta (3)-58304\right)
C_{A}^{2}T_{F}N_{\!f}+\left( 27648\zeta (3)+320\right)
C_{A}C_{F}T_{F}N_{\!f}\right.  \nonumber \\
&&\left.
~~~~~-~9728C_{A}T_{F}^{2}N_{\!f}^{2}-2304C_{F}^{2}T_{F}N_{%
\!f}-5632C_{F}T_{F}^{2}N_{\!f}^{2}\right] \frac{a^{3}}{1152}~+~O(a^{4}) \\
\gamma _{c}(a) &=&(\alpha -3)C_{A}\frac{a}{4}~+~\left[ \left( 3\alpha
^{2}-3\alpha -95\right) C_{A}^{2}+40C_{A}T_{F}N_{\!f}\right] \frac{a^{2}}{48}
\nonumber  \label{gc} \\
&&+~\left[ \left( 162\alpha ^{3}+1485\alpha ^{2}+(3672-2592\zeta (3))\alpha
-(1944\zeta (3)+63268)\right) C_{A}^{3}\right.  \nonumber \\
&&\left. ~~~~~-~\left( 6048\alpha -62208\zeta (3)-6208\right)
C_{A}^{2}T_{F}N_{\!f}-\left( 82944\zeta (3)-77760\right)
C_{A}C_{F}T_{F}N_{\!f}\right.  \nonumber \\
&&\left. ~~~~~+~9216C_{A}T_{F}^{2}N_{\!f}^{2}\right] \frac{a^{3}}{6912}%
~+~O(a^{4})\;, \\
\gamma_{\mathcal{O}}(a) &=& -~ \left[ 16T_{F}N_{\!f}+(3\alpha
-35)C_{A}\right] \frac{a}{6}  \nonumber \\
&&-~\left[ 280C_{A}T_{F}N_{\!f}+(3\alpha ^{2}+33\alpha
-449)C_{A}^{2}+192C_{F}T_{F}N_{\!f}\right] \frac{a^{2}}{24}  \nonumber \\
&&-~\left[ \left( (2592\alpha +1944)\zeta (3)+162\alpha ^{3}+2727\alpha
^{2}+18036\alpha -302428\right) C_{A}^{3}\right.  \nonumber \\
&&\left. ~~~~~-~\left( 62208\zeta (3)+6912\alpha -356032\right)
C_{A}^{2}T_{F}N_{\!f}+\left( 82944\zeta (3)+79680\right)
C_{A}C_{F}T_{F}N_{\!f}\right.  \nonumber \\
&&\left.
~~~~~-~49408C_{A}T_{F}^{2}N_{\!f}^{2}-13824C_{F}^{2}T_{F}N_{%
\!f}-33792C_{F}T_{F}^{2}N_{\!f}^{2}\right] \frac{a^{3}}{3456}~+~O(a^{4})~.
\label{gammao} \\
\gamma_{\alpha }(a) &=&\alpha \left[ \frac{a}{4}C_{A}~+~\left( \alpha
+5\right) C_{A}^{2}\frac{a^{2}}{16}\right.  \nonumber  \label{ga} \\
&&\left. ~~~~+~3C_{A}^{2}\left[ \left( \alpha ^{2}+13\alpha +67\alpha
\right) C_{A}-40T_{F}N_{\!f}\right] \frac{a^{3}}{128}\right] ~+~O(a^{4})
\end{eqnarray}
where the anomalous dimension of $\mathcal{O}$ in our conventions is given
by $(-4)$ times the result quoted in \cite{gracey}. The group Casimirs are $%
\mbox{tr}\left( T^a T^b \right)$~$=$~$T_F \delta^{ab}$, $T^a T^a$~$=$~$C_F I$%
, $f^{acd} f^{bcd}$~$=$~$\delta^{ab} C_A$, $N_{\! f}$ is the number of quark
flavours and $\zeta(n)$ is the Riemann zeta function. Our definition here of
$\gamma_\alpha(a)$, which denotes the running of $\alpha$, differs from that
of \cite{gracey} due to a different definition of $Z_\alpha$. For
computational reasons, it turns out to be more convenient to consider the
renormalization of the ghost operators $f^{abc}c^{b}c^{c}$, $f^{abc}%
\overline{c}^{b}c^{c}$ and $f^{abc}\overline{c}^{b}\overline{c}^{c}$ instead
of $gf^{abc}c^{b}c^{c}$, $gf^{abc}\overline{c}^{b}c^{c}$ and $gf^{abc}%
\overline{c}^{b}\overline{c}^{c}$ respectively. We note that we have first
verified that to three loops the anomalous dimension of each of the three
operators are in fact equal, in agreement with \cite{Dudal:2003dp}.
Accordingly, we find
\begin{eqnarray}
\gamma_{c^2}(a) &=& \frac{3}{2} C_A a ~+~ \left[ \left( 18\alpha + 95
\right) C_A^2 - 40 C_A T_F N_{\! f} \right] \frac{a^{2}}{24}  \nonumber \\
&& +~ \left[ \left( 621\alpha^2 + ( 7182 + 2592\zeta(3) )\alpha + (
1944\zeta(3) + 63268 ) \right) C_A^3 \right.  \nonumber \\
&& \left. ~~~~~-~ \left( 432\alpha + 62208\zeta(3) + 6208 \right) C_A^2 T_F
N_{\! f} + \left( 82944\zeta(3) - 77760 \right) C_A C_F T_F N_{\! f} \right.
\nonumber \\
&& \left. ~~~~~-~ 8960 C_A T_F^2 N_{\! f}^2 \right] \frac{a^3}{3456} ~+~
O(a^4) ~.
\end{eqnarray}
We have deduced this result using the \textsc{Mincer} package, \cite{mincer}%
, written in \textsc{Form}, \cite{form}, where the Feynman diagrams are
generated in \textsc{Form} input format by \textsc{Qgraf}, \cite{qgraf}. For
instance, for $f^{abc} \bar{c}^b c^c$ there are $529$ diagrams to determine
at three loops and $376$ for the operator $f^{abc} c^b c^c$ where each is
inserted in the appropriate ghost two-point function. The same \textsc{Form}
converter functions of \cite{gracey} were used here. Since the operator $%
f^{abc} \bar{c}^b c^c$ has the same ghost structure as the operator $\bar{c}%
^a c^a$, we were able merely to replace the Feynman rule for the operator
insertion of $\bar{c}^a c^a$ in the ghost two-point function with the new
operator and use the same routine which determined $\gamma_{\mathcal{O}}(a)$
in \cite{gracey}. However, as $f^{abc} c^b c^c$ has a different structure we
had to generate a new \textsc{Qgraf} set of diagrams to renormalize this
operator. That the anomalous dimensions of both operators emerged as
equivalent at three loops for all $\alpha$ provides a strong check on our
programming as well as justifying the general result of section 2. Now,
taking into account the extra factor $g$, the anomalous dimension $%
\gamma_{gc^{2}}(a)$ is found to be
\begin{eqnarray}
\gamma_{gc^2}(a) &=& \left[ 8 T_F N_{\! f} - 13 C_A \right] \frac{a}{6} ~+~
\left[ \left( 6\alpha - 59 \right) C_A^2 + 40 C_A T_F N_{\! f} + 32 C_F T_F
N_{\! f} \right] \frac{a^2}{8}  \nonumber \\
&& +~ \left[ \left( 207\alpha^2 + ( 2394 + 864\zeta(3) )\alpha + (
648\zeta(3) - 39860 ) \right) C_A^3 \right.  \nonumber \\
&& \left. ~~~~~-~ \left( 144\alpha + 20736\zeta(3) - 58304 \right) C_A^2 T_F
N_{\! f} + \left( 27648\zeta(3) + 320 \right) C_A C_F T_F N_{\! f} \right.
\nonumber \\
&& \left. ~~~~~-~ 9728 C_A T_F^2 N_{\! f}^2 - 2304 C_F^2 T_F N_{\! f} - 5632
C_F T_F^2 N_{\! f}^2 \right] \frac{a^3}{1152} ~+~ O(a^4) ~.  \label{ggc2}
\end{eqnarray}
It is then easily checked from the expressions (\ref{gammao}), (\ref{ga})
and (\ref{ggc2}) that, up to three loop order,
\begin{equation}
\gamma_{\mathcal{O}}(a) ~=~ -~ 2\left( \gamma _{c}(a) ~+~ \gamma
_{gc^{2}}(a) \right) ~.  \label{gjveri}
\end{equation}
It is worth mentioning that the renormalizability of the operator $\mathcal{O%
}$ was already discussed in \cite{deBoer:1995dh} from the viewpoint of the
massive Curci-Ferrari model. Whilst the relation (\ref{gj}) was not
explicitly given in \cite{deBoer:1995dh}, it is possible to obtain the
relation from that analysis. Although the relation (\ref{gj}) has been
established in the case of the Curci-Ferrari gauge, it expresses a general
property of the gluon-ghost operator which remains valid also in the Landau
and maximal abelian gauges, as will be shown in the following sections.

\section{The Landau gauge}

The Landau gauge is a particular case of the Curci-Ferrari gauge,
corresponding to $\alpha =0$. The Landau gauge is known to possess further
additional Ward identities \cite{book,lg}, implying that the renormalization
constants $Z_{L}$ and $Z_{c}$ can be expressed in terms of $Z_{g}$ and $%
Z_{A} $, according to \cite{dsv}
\begin{equation}
\,\,Z_{L}=\,Z_{A}^{1/2}\,\,\,\,,\,\,\,\,Z_{c} ~=~ Z_{g}^{-1}Z_{A}^{-1/2}.
\label{zlcl}
\end{equation}
Therefore, it follows that (\ref{zj}) reduces to
\begin{equation}
Z_{J} ~=~ Z_{g}Z_{A}^{-1/2}  \label{zjl}
\end{equation}
from which the expression (\ref{ada2}) is recovered, providing a nontrivial
check of the validity of the general relationship (\ref{go}).

As another internal check of our computations, we note that we should also
find in the Landau gauge that $\gamma_{gc^{2}}(a)$ $=$ $\gamma_{A}(a)$, as
is obvious from (\ref{zlcl}). It can indeed be checked from (\ref{gA}) and (%
\ref{ggc2}) that
\begin{equation}  \label{check}
\left.\gamma_{gc^{2}}(a)\right|_{\alpha=0} ~=~
\left.\gamma_{A}(a)\right|_{\alpha=0} ~.
\end{equation}

\section{The maximal abelian gauge}

As is well known, the maximal abelian gauge is a nonlinear partial gauge
fixing allowing for a residual $U(1)^{N-1}$ local invariance \cite
{ell,mag1,fz,mag2}. In the following, a Landau type gauge fixing will be
assumed for this local residual invariance. The Slavnov-Taylor and the $%
\delta $ Ward identities (\ref{sti}) and (\ref{wwi}) can be
straightforwardly generalized to this case. It is useful to recall that the
gauge field is now decomposed into its off-diagonal and diagonal components
\begin{equation}
A_{\mu }^{a}T^{a}=A_{\mu }^{i}T^{\,i}+A_{\mu }^{\alpha }T^{\alpha }\;,
\label{conn-sun}
\end{equation}
where the index $i$ labels the $N-1$ generators $T^{\,i}$ of the Cartan
subalgebra of $SU(N)$. The remaining $N(N-1)$ off-diagonal generators $%
T^{\alpha }$ will be labelled by the index $\alpha $. Accordingly, for the
Faddeev-Popov ghost $c^{a}$ we have
\begin{equation}
c^{a}T^{a}=c^{i}T^{\,i}+c^{\alpha }T^{\alpha }\;,  \label{ghd}
\end{equation}
with
\begin{eqnarray}
sc^{\alpha } &=&gf\,^{\alpha \beta i}c^{\beta }c^{i}+\frac{g}{2}f\,^{\alpha
\beta \gamma }c^{\beta }c^{\gamma
},\,\,\,\,\,\,\,\,\,\,\,\,\,\,\,\,\,\,\,\,\,\,  \label{ghs} \\
\,\,\,\,\,\,\,\,\,\,\,\,\,\,\,\,\,sc^{i} &=&\frac{g}{2}\,f\,^{i\alpha \beta
}c^{\alpha }c^{\beta }.  \nonumber
\end{eqnarray}
Also, the group index of the gluon-ghost operator runs only over the
off-diagonal components, namely
\begin{equation}
\mathcal{O}_{\mathrm{{\mathrm{MAG}}}} ~=~ \left( \frac{1}{2}A^{\alpha\mu}
A_{\mu}^{\alpha }+\alpha \overline{c}^{\alpha }c^{\alpha }\right) \;.
\label{omag}
\end{equation}
Denoting respectively by $\widetilde{Z}$ and $Z$ the renormalization factors
of the off-diagonal and diagonal components of the fields, it follows that,
according to the relationship (\ref{gj}), the output of the Slavnov-Taylor
and $\delta $ Ward identities gives
\begin{equation}
\gamma_{\mathcal{O}_{\mathrm{{\mathrm{MAG}}}}}(a) ~=~ -~ 2\left( \widetilde{%
\gamma}_{c^{\alpha}}(a) +\widetilde{\gamma }_{gc^{2}}(a) \right) \;,
\label{gomag}
\end{equation}
where $\widetilde{\gamma}_{c^{\alpha }}(a)$ and $\widetilde{\gamma }%
_{gc^{2}}(a)$ are the anomalous dimensions of the off-diagonal ghost $%
c^{\alpha }$ and of the composite operator $\left( gf\,^{\alpha \beta
i}c^{\beta }c^{i}+\frac{g}{2}f\,^{\alpha \beta \gamma }c^{\beta }c^{\gamma
}\right)$ which corresponds to the BRST variation of $c^{\alpha }$.
Moreover, as shown in \cite{fz}, the use of the Landau gauge for the local
residual $U(1)^{N-1}$ invariance$\;$allows for a further Ward identity. This
identity, called the diagonal ghost Ward identity in \cite{fz}, implies that
the anomalous dimension $\widetilde{\gamma }_{gc^{2}}(a)$ can be expressed
as
\begin{equation}
\widetilde{\gamma }_{gc^{2}}(a) ~=~ \frac{\beta (a)}{a} ~-~ \widetilde{%
\gamma }_{c^{\alpha }}(a) ~-~ \gamma _{c^{i}}(a) \;,  \label{rel}
\end{equation}
where $\gamma _{c^{i}}(a)$ is the anomalous dimension of the diagonal ghost $%
c^{i}$. Therefore, for the expression of $\gamma_{\mathcal{O}_{\mathrm{{%
\mathrm{MAG}}}}}(a)$ we obtain
\begin{equation}
\gamma _{\mathcal{O}_{\mathrm{{\mathrm{MAG}}}}}(a) ~=~ -~ 2\left( \frac{%
\beta(a)}{a} ~-~ \gamma_{c^{i}}(a) \right) \;,  \label{gomagf}
\end{equation}
a result which is in complete agreement with that already found in \cite{ell}
for the case of $SU(2)$. Finally, it is worth mentioning that the anomalous
dimensions of the diagonal and off-diagonal components of the fields have
been computed at one-loop order in \cite{ell,mag2}, so that (\ref{gj}) gives
explicit knowledge of the one-loop anomalous dimension of the gluon-ghost
operator in the maximal abelian gauge.

\section{Conclusion}

We have shown that the mass dimension two gluon-ghost operator $\mathcal{O}=%
\frac{1}{2}A_{\mu }^{a}A^{\mu a}+\alpha \overline{c}^{a}c^{a}$ is
multiplicatively renormalizable in the Landau, Curci-Ferrari and maximal
abelian gauges. Further, we were able to establish a general relation
between the anomalous dimension of $\mathcal{O}$, the Faddeev-Popov ghost $%
c^{a}$ and the dimension two ghost operator $gf^{abc}c^{b}c^{c}$, as
expressed by the eq.(\ref{go}). This relation has been derived within the
framework of the algebraic renormalization \cite{book}, following from the
Slavnov-Taylor identity (\ref{sti}). As such, it extends to all orders of
perturbation theory and is renormalization scheme independent, for any
scheme preserving the Slavnov-Taylor identity. It has been explicitly
verified up to three loops in the $\overline{\mathrm{MS}}$ scheme in the
Curci-Ferrari gauge.

Furthermore, due to additional Ward identities that exist in the
Landau gauge \cite{book} and in the MAG \cite{fz}, we were able to
rewrite the relation (\ref{go}) for the anomalous dimension for
$\mathcal{O}$ in terms of the $\beta $-function and the anomalous
dimension of the gluon and/or ghost fields. In particular,
concerning the maximal abelian gauge, it is worth underlining that
the multiplicative renormalizability of the gluon-ghost operator,
eq.(\ref{gomagf}), is a necessary ingredient towards the
construction of a renormalizable effective potential for studying
the possible condensation of the gluon-ghost operator and the
ensuing dynamical mass generation, as done in the Landau
\cite{vland} and Curci-Ferrari \cite {cfep} gauges.

As a final remark, we point out that, from the 3-loop expressions
given in section 3, it is easily checked that the following
relations holds in the Curci-Ferrari gauge:
\begin{equation}
\gamma _{\mathcal{O}}(a)~=~-~\left( \frac{\beta (a)}{a}~+~\gamma
_{A}(a)\right) ,  \label{r1}
\end{equation}
\begin{equation}
\gamma _{gc^{2}}(a)~=~\gamma _{A}(a)~-~2\gamma _{\alpha }(a)~.  \label{r2}
\end{equation}
Up to now, we do not know if these relations are valid to all
orders. They do not follow from the Slavnov-Taylor identity
(\ref{sti}). Nevertheless, although eqs.(\ref{r1}), (\ref{r2})
have been obtained in a particular renormalization scheme, i.e.
the $\overline{\mathrm{MS}}$ scheme, it could be interesting to
search for additional Ward identities in the Curci-Ferrari gauge
which, as in the case of the Landau gauge \cite{dsv}, could allow
for a purely algebraic proof of eqs.(\ref{r1}), (\ref{r2}). Notice
in fact that, when $\alpha =0$, eq.(\ref{r1}) yields the anomalous
dimension of the composite operator $A^2_\mu$ in the Landau gauge.
Also, eq.(\ref{r2}) reduces to the relation (\ref{check}) of the
Landau gauge, since $\gamma _{\alpha }(a)$ $\equiv $ $0$ if
$\alpha $ $=$ $0$.

\vspace{1cm}

\textbf{Acknowledgments}. M.P. is grateful to R. Ferrari for useful
discussions. We acknowledge the Conselho Nacional de Desenvolvimento
Cient\'{i}fico e Tecnol\'{o}gico (CNPq-Brazil), the Funda{\c{c}}{\~{a}}o de
Amparo a Pesquisa do Estado do Rio de Janeiro (Faperj), the SR2-UERJ, the
Coordena{\c{c}}{\~{a}}o de Aperfei{\c{c}}oamento de Pessoal de N\'{i}vel
Superior (CAPES) and the MIUR-Italy for financial support. D.D. would like
to thank the kind hospitality of the Theoretical Physics Department of the
UERJ where part of this work was completed.


\begin{thebibliography}{99}
\bibitem{bou}  P.~Boucaud, G.~Burgio, F.~Di Renzo, J.~P.~Leroy, J.~Micheli,
C.~ Parrinello, O.~P$\grave{\mathrm{e}}$ne, C.~Pittori,
J.~Rodriguez-Quintero, C.~Roiesnel and K.~Sharkey, \emph{JHEP} \textbf{0004}
(2000) 006;

P.~Boucaud, A.~Le Yaouanc, J.~P.~Leroy, J.~Micheli, O.~P$\grave{\mathrm{e}}$%
ne and J.~Rodriguez-Quintero, \emph{Phys.\ Rev.}\ \textbf{D63} (2001) 114003;

P.~Boucaud, J.~P.~Leroy, A.~Le~Yaouanc, J.~Micheli, O.~P\`{e}ne, F.
~De~Soto, A.~Donini, H.~Moutarde and J.~Rodriguez-Quintero, \emph{Phys.\ Rev.%
}\ \textbf{D66} (2002) 034504;

Ph. Boucaud, F. De Soto, A. Le Yaouanc, J.P. Leroy, J. Micheli, H. Moutarde,
O. P$\grave{\mathrm{e}}$ne and J.~Rodriguez-Quintero, \emph{Phys.\ Rev.}\
\textbf{D67} (2003) 074027.

\bibitem{gz}  F.~V.~Gubarev, L.~Stodolsky and V.~I.~Zakharov, \emph{Phys.\
Rev.\ Lett.}\ \textbf{86} (2001) 2220;

F.~V.~Gubarev and V.~I.~Zakharov, \emph{Phys.\ Lett.}\ \textbf{B501} (2001)
28.

\bibitem{vland}  H.~Verschelde, K.~Knecht, K.~Van Acoleyen and
M.~Vanderkelen, \emph{Phys.\ Lett.}\ \textbf{B516} (2001) 307,
Erratum-ibid., to appear.

\bibitem{ope}  K.-I. Kondo, \emph{Phys. Lett.} \textbf{B514} (2001) 335.

K.-I. Kondo, T. Murakami, T. Shinohara \& T. Imai, \emph{Phys. Rev.} \textbf{%
D65} (2002) 085034.

\bibitem{cfep}  D.~Dudal, H.~Verschelde, V.~E.~Lemes, M.~S.~Sarandy,
S.~P.~Sorella and M. Picariello, \emph{Gluon-ghost condensate of mass
dimension 2 in the Curci-Ferrari gauge}, \emph{Ann. Phys.}, to appear,
hep-th/0302168.

\bibitem{book}  O.~Piguet and S.~P.~Sorella, \emph{Algebraic Renormalization}%
, Monograph series \textbf{m28}, Springer Verlag, 1995

\bibitem{vlco}  K.~Knecht and H.~Verschelde, \emph{Phys.\ Rev.}\ \textbf{D64}
(2001) 085006.

\bibitem{gracey}  J.~A.~Gracey, \emph{Phys.\ Lett.}\ \textbf{B552} (2003)
101.

\bibitem{dsv}  D. Dudal, H. Verschelde, and S.P. Sorella, \emph{Phys.\ Lett.}%
\ \textbf{B555} (2003) 126.

\bibitem{ell}  U.~Ellwanger and N.~Wschebor, \emph{Int. J. Mod. Phys. }%
\textbf{A18 }(2003) 1595

\bibitem{obt}  I. Ojima, \emph{Z. Phys.}\textbf{\ C13} (1982) 173;

R. Delbourgo and P.D. Jarvis, \emph{J. Phys. }\textbf{A15} (1982) 611;

L. Baulieu and J. Thierry-Mieg, \emph{Nucl. Phys.}\textbf{\ B197} (1982) 477.

\bibitem{cf}  G. Curci and R. Ferrari, \emph{Nuovo Cim.} \textbf{A32} (1976)
151;

G. Curci and R. Ferrari, \emph{Phys.\ Lett.}\ \textbf{B63} (1976) 91.

\bibitem{fd}  F. Delduc and S.P. Sorella, \emph{Phys.\ Lett.}\ \textbf{B231}
(1989) 408.

\bibitem{sl2r}  D. Dudal, V.E.R. Lemes, M. Picariello, M.S. Sarandy, S.P.
Sorella, and H. Verschelde, \emph{JHEP} \textbf{0212} (2002) 008.

\bibitem{Dudal:2003dp}  D.~Dudal, H.~Verschelde, V.~E.~Lemes, M.~S.~Sarandy,
S.~P.~Sorella, M.~Picariello, A. Vicini, J. A. Gracey, \emph{JHEP} \textbf{%
0306} (2003) 003.

\bibitem{mincer}  K.G. Chetyrkin, A.L. Kataev and F.V. Tkachov, \emph{Nucl.
Phys.} \textbf{B174} (1980) 345.

\bibitem{form}  J.A.M. Vermaseren, math-ph/0010025.

\bibitem{qgraf}  P. Nogueira, \emph{J. Comput. Phys.} \textbf{105} (1993)
279.

\bibitem{deBoer:1995dh}  J.~de Boer, K.~Skenderis, P.~van Nieuwenhuizen and
A.~Waldron, \emph{Phys.\ Lett.} \textbf{B367} (1996) 175

\bibitem{lg}  A. Blasi, O.\ Piguet, S.P. Sorella, \emph{Nucl. Phys.}\textbf{%
\ B356} (1991) 154.

\bibitem{mag1}  H. Min, T. Lee and P.Y. Pac, \emph{Phys. Rev.} \textbf{D32}
(1985) 440.

\bibitem{fz}  A.R. Fazio, V.E.R. Lemes, M.S. Sarandy and S.P. Sorella, \emph{%
Phys. Rev. }\textbf{D64} (2001) 085003.

\bibitem{mag2}  T. Shinohara, T. Imai, K.-I. Kondo, hep-th/0105268.
\end{thebibliography}
\end{document}